\documentclass[twocolumn,aps,prb,a4paper,reprint,unsortedaddress] {revtex4} 
\usepackage{graphicx}
\usepackage{amsmath}
\newcommand{\brahket}[3]{\langle #1 | #2 | #3 \rangle}
\newcommand{\ket}[1]{| #1 \rangle}

\begin{document}
\pagestyle{empty} 

\title{Vibrationally induced flip motion of a hydroxyl dimer on Cu(110)}

\date{\today}

\author{Yasuhiro Ootsuka}
\affiliation{Division of Nano and New Functional Material Science, 
Graduate School of Science and
Engineering, University of Toyama, Toyama, 930-8555 Japan}

\author{Thomas Frederiksen}
\affiliation{Donostia International Physics Center (DIPC), ES-20018 San Sebasti\'an, Spain}

\author{Hiromu Ueba}
\affiliation{Division of Nano and New Functional Material Science, 
Graduate School of Science and
Engineering, University of Toyama, Toyama, 930-8555 Japan}

\author{Magnus Paulsson}
\affiliation{School of Computer Science, Physics and Mathematics, Linnaeus University, SE-391 82 Kalmar, Sweden}
\email{magnus.paulsson@lnu.se}

\begin{abstract}
Recent low-temperature scanning-tunneling microscopy experiments 
[T. Kumagai {\it et al.}, Phys. Rev. B {\bf 79}, 035423 (2009)] 
observed the vibrationally induced flip motion 
of a hydroxyl dimer (OD)$_2$ on Cu(110). 
We propose a model to describe two-level fluctuations and current-voltage characteristics
of nanoscale systems which undergo vibrationally induced switching. The parameters of the 
model are based on comprehensive density-functional calculations of the system's 
vibrational properties. 
For the dimer (OD)$_2$ the calculated population of the high and low conductance 
states, the $I-V$, $dI/dV$, and $d^2I/dV^2$ curves are in good agreement with the 
experimental results and underlines the different roles played by the 
free and shared OD stretch modes of the dimer. 
\end{abstract}

\pacs{}
\keywords{}

\maketitle

Electron transport through single-molecule junctions has been receiving enthusiastic 
interest for a development of novel molecular devices. Nonlinear $I-V$ characteristics 
associated with the vibrationally mediated configurational change with different conductances
have been observed in a series of systems such as pyrrolidine on a Cu(001)~\cite{Gaudioso}, H$_2$ on Cu~\cite{Gupta},
CO bridging a Pt contact~\cite{Thijssen}, and H$_2$ in Au contacts\cite{AuH2}. 
In these systems $dI/dV$ spectra show
anomalous spikes---in contrast to steps usually observed in inelastic electron tunneling 
spectroscopy (IETS)\cite{Stipe_Science}---at the bias voltage related to the vibrational 
mode energies.

Recently Kumagai {\it et al.}\cite{Kumagai_2009} studied the dynamics of a single hydroxyl (OH, OD) 
molecule and the dimer (OD)$_2$ on Cu(110) using a scanning tunneling 
microscope (STM). The STM images observed for the monomer 
suggested the possible quantum tunneling of a hydroxyl between two equivalent 
adsorption configurations on Cu(110), as supported by the density functional 
theory (DFT) calculations of the transition path and rate of the flipping of 
OH on Cu(110)~\cite{Davidson}. This spontaneous flip motion of hydrogen atoms 
in the monomer is quenched for the dimer at low temperatures, but can be induced by excitation 
of the OH/OD stretch mode by tunneling electrons. Time-averaged measurements of 
the current show a non-linear current ($I$) increase at the bias voltage ($V$) 
inducing the transition from the high and low conductance states. The 
appearance of the peak in $dI/dV$ and the peak and dip in $d^2I/dV^2$  
from transitions between states with distinct conductances have also been 
reported previously~\cite{Gupta,Thijssen,AuH2}.

In this paper a combined use of the DFT-based SIESTA~\cite{SIESTA}, 
TranSIESTA~\cite{TRANSIESTA} and Inelastica~\cite{INELASTICA,INELASTICA2} packages 
permits us to gain insight into the elementary processes that induce 
the flip motion of the asymmetric dimer. 
The extensive DFT calculations provide the ground state geometry, 
vibrational modes, electron-vibration couplings, emission rate of 
vibrations from tunneling electrons, vibrational damping due to 
electron-hole pair excitation, and the high and low conductance. 
These calculated properties allow us to model the population of the 
high and low conductance states 
as a function of the bias voltage and the nonlinear $I-V$ characteristics 
for (OD)$_2$ on a Cu(110) surface. The experimental results 
(relative occupation of the high and low conductance 
state, $I-V$ curve, and $dI/dV$) are nicely reproduced, and 
the different roles played by the free and shared OD stretch 
modes in the vibrationally mediated configurational flip motion are clarified. 

\begin{figure}
\begin{center}
\includegraphics[width=\columnwidth]{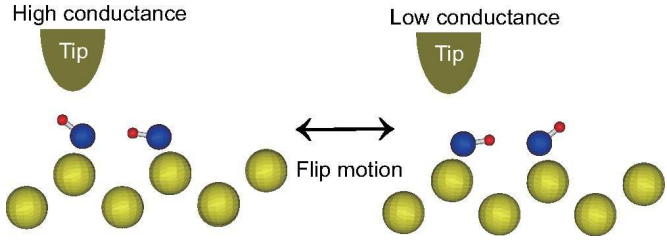}
\caption{(Color online)
 Schematic view of the high and low conductance states of the dimer (OD)$_2$ on Cu(110). 
Yellow, blue and red circles are Cu, O, and D atoms, respectively. 
See text for details on the relaxed geometry.
\label{fig.1}
} 
\end{center}
\end{figure}

The telegraph switching between high- and low-conductance states of 
(OD)$_2$ is shown schematically in Fig.~\ref{fig.1}. In low-temperature 
STM, no spontaneous switching is observed at low bias since the barrier for the 
reaction is substantial\cite{Kumagai_2009}. 
Keeping the STM tip stationary and increasing bias over 
approximately 200 meV the switching between the two degenerate low-energy
configurations is triggered by phonon emission. However, the reaction rate
remains rather small until the bias exceeds the $\nu$(OD) stretch vibrations ($\approx$ 300-330 meV). We denote the 
two configurations as the high ($H$) and low ($L$) conductance states where the tip-dimer distance is 
smaller for the $H$ configuration. 

To investigate the flip motion of the dimer (OD)$_2$ we performed calculations on a 6 atom thick Cu $4\times3$ 
slab using periodic boundary conditions. The tip was modeled as one 
protruding Cu atom on the reverse side of the slab. All calculations were 
performed with the PBE GGA functional, DZP (SZP) basis set for dimer (bulk Cu), 
$3\times4$ $k$-point sampling, and a 300 Ry mesh cutoff. 
The elastic transport properties were calculated using the DFT+NEGF method 
where 6 additional layers of Cu were used to connect the central region to 
semi-infinite metallic leads. 
Fig.~\ref{fig.1} shows the schematic side view of the 
dimer in the $H$ and $L$ conductance states.
The calculated low-bias conductance
ratio $G_H/G_L=2.0$ closely matches the experimental ratio. However, the calculated 
absolute conductances are much larger than in the experiment since numerical considerations 
demand the calculations to be performed at a small tip-dimer distance.
The relaxed hydrogen-bonded OD-O distance is 2.90 {\AA} while the OD bond length and
tilt angle to the surface normal are
1.00 (0.98) {\AA} and 81$^\circ$ (51$^\circ$), respectively. 
The two hydroxyl groups, bonded near adjacent 
bridge sites along (001) are oriented according to
the optimal configuration for hydrogen bonding. The D atom pointing towards the 
adjacent OD molecule 
form the OD-O bond (with characteristic shared OD stretch mode) in contrast to the D atom 
pointing away from the adjacent OD molecule (with free OD stretch mode). These 
nonequivalent configurations of each OD molecule lift the degeneracies of 
the vibrational energies for the two OD molecules.

\begin{figure}
\begin{center}
\includegraphics[width=0.45\textwidth]{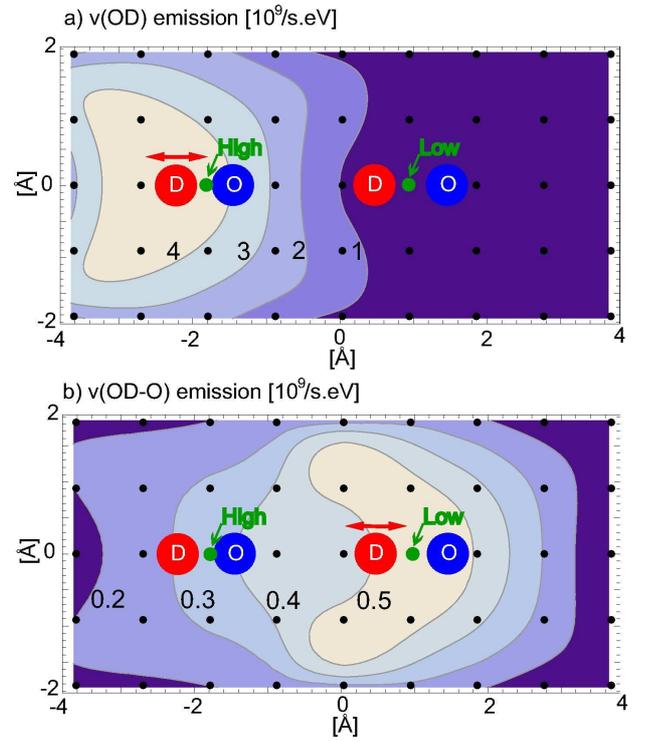}
\caption{(Color online) (a) Emission rate constant $\lambda_{\mathrm{em},\nu(\mbox{OD})}$ of the free 
$\nu$(OD) stretch and (b) $\lambda_{\mathrm{em},\nu(\mbox{OD-O})}$ of the 
shared $\nu$(OD-O) stretch mode. The blue and red circle are O and D atom, respectively. 
The black doted grids are the tip position fixed at 5.6 {\AA} above the oxygen atom and
the red arrows indicate the vibrational motion. 
The values of the contour lines are indicated in the figures [$10^9$/(s eV)] and the exact values 
with the tip positioned at the large green dots are listed in Table~\ref{tab.hw}.
\label{fig.emissionmap}
} 
\end{center}
\end{figure}

The vibrational modes and frequencies of the molecular adsorbates and the corresponding electron-phonon couplings are 
calculated from a finite difference scheme\cite{INELASTICA,INELASTICA2}. 
These calculations were repeated with the tip scanned over the surface as indicated by the black dots in Fig.~\ref{fig.emissionmap}.
Table I lists the calculated 
vibrational modes for the $H$ and $L$ configurations. Here $\nu$(OD) 
labels the free OD stretch mode, $\nu$(OD-O) the shared OD stretch mode involving the D atom
between two oxygen atoms, and rot$_{xy(z)}$ rotation 
modes in the surface plane (surface normal). Values for the 
low-energy modes mainly involving motion of the oxygen atoms have been omitted 
from the table.

In addition to the vibrational energies ($\hbar\omega$), the electron-hole pair damping rate 
$\gamma_{{eh}}$ and vibration generation rate $\gamma_{\mathrm{em}}$ were calculated for each tip position 
within the lowest order expansion (LOE) scheme\cite{INELASTICA,INELASTICA2}. 
In the low-temperature limit the emission rate is give by
\begin{equation}
\gamma_{\mathrm{em}}(V) = \lambda_{\mathrm{em}} \left( \left|eV\right|-\hbar \omega \right) \theta \left(  \left|eV\right|-\hbar \omega \right)
\label{eq.emission}
\end{equation}
where $V$ is the applied bias, $\theta$ the step function, and 
$\lambda_{\mathrm{em}}=\frac{1}{\pi\hbar}\mathrm{Tr}[\mathbf M \mathbf A_1 \mathbf M \mathbf A_2]=
\frac{4 \pi}{\hbar} \sum_{\alpha \beta} \left| \brahket{\psi_\beta^2}{\mathbf M}{\psi_\alpha^1}\right|^2$ 
the emission rate constant written in terms of electron-phonon coupling matrix $\mathbf M$ and 
partial spectral density $\mathbf A_{1/2}$ at the Fermi energy, cf.~Ref.~\onlinecite{INELASTICA}. 
The spectral density can further be rewritten in terms of a sum over scattering states $\ket{\psi_\alpha^{1/2}}$ 
at the Fermi-energy incident from substrate / tip (1/2) showing the equivalence of the LOE scheme with 
the Fermi-golden rule, assuming a constant density of states around the Fermi-energy.\cite{INELASTICA} 
Furthermore, for weak coupling between tip-adsorbate, the emission rate 
constants scale with the tip-adsorbate coupling ($\tau$) squared ($\lambda \propto \tau^2$). Since the current scales 
in the same way, the emission rate at a given voltage is proportional to the current.
The electron-hole pair damping rate $\gamma_{{eh}}$ is insensitive to the 
position of the STM tip since the damping is dominated by the metal surface.
The emission rate map for the two $\nu$(OD) and $\nu$(OD-O) modes are shown in 
Fig.~\ref{fig.emissionmap} where the emission rates were calculated by scanning the tip over the surface 
at a constant height of  5.6 \AA \ above the oxygen atoms, i.e., the DFT calculations of e-ph coupling, 
transmission, emission rates, and damping were repeated for geometries with the tip displaced relative to the OD dimer.
We note that the emission rate for the free OD stretch is more localized around 
the high conductance site than the shared OD-O stretch mode which is also evident 
from Table I. The green points in Fig.~\ref{fig.emissionmap} correspond to the 
$H$ and $L$ configurations listed in Table~\ref{tab.hw}.

\begin{table}
\caption{Vibrational mode frequencies, emission rate constants and their electron-hole pair damping rates at $H$ and $L$ conductance state.
\label{tab.hw}
}
\begin{ruledtabular}
\begin{tabular}{lrrr}
Mode & $\hbar\omega \left[{\rm meV}\right]$ &  $\lambda_{\mathrm{em}} \left[10^9/({\rm s\;eV})\right]$ & $\gamma_{{eh}}\left[10^9/{\rm s}\right]$  \\
\hline
$\nu$(OD) $L$ &327.2& 0.17 & 47.3 \\
$\nu$(OD) $H$ & 326.6 & 4.42 & 52.3\\
$\nu$(OD-O) $L$ &301.2& 0.49 & 248.0 \\
$\nu$(OD-O) $H$ & 301.8 & 0.35 & 249.6\\
rot$_z$(OD-O) $L$ &76.8& 1.19 & 250.7 \\
rot$_z$(OD-O) $H$ & 76.5 & 0.67 & 268.0\\
\hline
rot$_{xy}$(OD-O) $L$ & 77.0 & 0.24 & 103.7\\
rot$_{xy}$(OD-O) $H$ & 76.9 & 0.10 & 76.0\\
rot$_{xy}$(OD) $L$ & 51.8 & 0.01 & 59.4\\
rot$_{xy}$(OD) $H$ & 52.6 & 0.71 & 60.7\\
rot$_{z}$(OD) $L$ & 49.9 & 0.27 & 97.0\\
rot$_{z}$(OD) $H$ & 49.8 & 0.79 & 86.6\\
\end{tabular}
\end{ruledtabular}
\end{table}

To describe vibrationally induced switching in nanoscale systems we propose a simple 
model which expresses the current in terms of the occupation $n_{H(L)}$ and 
conductance $\sigma_{H(L)}$ of the $H$ and $L$ conductance states, 
\begin{equation}
I=\sigma_{H}n_{H}(V)V+\sigma_{L}n_{L}(V)V,
\label{eq.IV}
\end{equation}
where $n_{H(L)}$ is determined as a stationary solution of the rate equations,
\begin{eqnarray}
dn_{H}/dt&=&\Gamma^{L\rightarrow H}(V)n_{L}-\Gamma^{H\rightarrow L}(V) n_{H}\\
dn_{L}/dt&=&\Gamma^{H\rightarrow L}(V)n_{H}-\Gamma^{L\rightarrow H}(V) n_{L}
\end{eqnarray}
with the condition $n_{H}+n_{L}=1$. This gives
\begin{equation}
n_{L}(V)=\frac{\Gamma^{H\rightarrow L}}{\Gamma^{H\rightarrow L}+\Gamma^{L\rightarrow H}}, \quad
n_{H}(V)=\frac{\Gamma^{L\rightarrow H}}{\Gamma^{H\rightarrow L}+\Gamma^{L\rightarrow H}},
\label{eq.ratesol}
\end{equation}
where $\Gamma^{H\rightarrow L}$ is the transition rate from 
$H$ to $L$.
The experiments show that the reaction rate at a constant voltage is propotional to the current, i.e., 
the reaction rate is proportional to the emission rate of the vibrations. 
The simplest way to model the transition rates is therefore to assume a linear dependence 
on the vibrational occupation $\propto \gamma_{\mathrm{em}}/\gamma_{\mathrm{eh}}$ or equivalently the vibrational generation rates $\gamma_{\mathrm{em}}$\cite{Ueba_Sergei_2010}, i.e.,
\begin{equation}
\Gamma^{H\rightarrow L(L\rightarrow H)}(V)  =  \Gamma^{H(L)}_0 +\sum_{i} C_{i} \Gamma^{H(L)}_{\mathrm{em},i}(V) ,
\label{eq.rate}
\end{equation}
where a constant rate $\Gamma^{H(L)}_0$ is introduced to model the collective effect of many low-energy vibration modes
and where $\Gamma^{H(L)}_{\mathrm{em},i}$ represents the broadened 
vibrational generation rate\cite{Motobayashi} of a distinct mode $i$ given by
\begin{equation}
\Gamma^{H(L)}_{\mathrm{em},i}=\int_0^\infty W(\omega-\Omega_i^{H(L)},\sigma_{\mathrm{ph}}^i) \gamma_{\mathrm{em},i}(eV,\omega) d\omega.
\end{equation}
In the above equation we use a Gaussian distribution function $W(\omega,\sigma_{\mathrm{ph}}) \propto e^{-\omega^2/(2 \sigma_\mathrm{ph}^2)}$ 
characterized by a standard deviation $\sigma_\mathrm{ph}$.
The prefactors $C_{i}$ describe the likelihood of the generated vibrational excitation to induce 
the flip motion, i.e., the probability of transferring the motion from the vibration to the reaction coordinate\cite{Ueba_Sergei_2010}. 

\begin{figure}
\begin{center}
\includegraphics[width=0.4\textwidth]{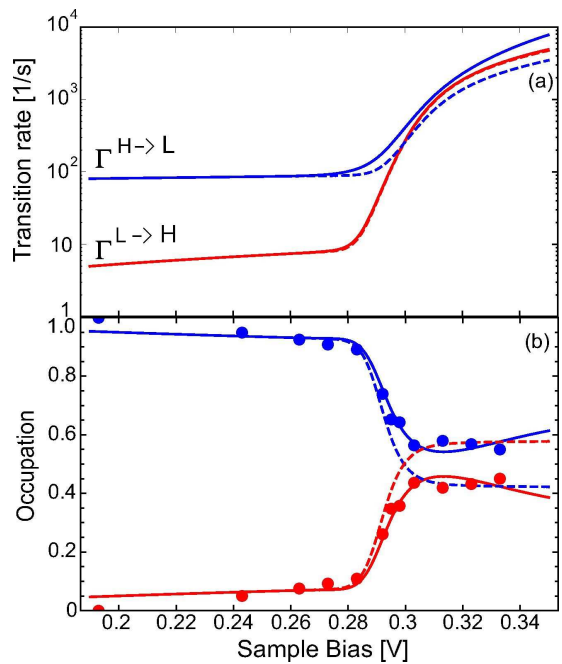}
\caption{(Color online) (a) Calculated transition rates $\Gamma^{H\rightarrow L(L\rightarrow H)}(V)$ and 
(b) occupations $H$ (red) and $L$ (blue) configurations of a dimer (OD)$_2$ on Cu(110). See text for the parameters 
used to fit the experimental occupations from Ref.~\onlinecite{Kumagai_2009} (blue/red dots).
The dashed curves are calculated without the contribution of the free OD stretch mode. 
\label{fig.occupation}
}
\end{center}
\end{figure}

We obtain a good fit to the experimental OD dimer data from Ref.~\onlinecite{Kumagai_2009}
considering only three different vibrational modes, see Fig. 
\ref{fig.occupation}. The fitting parameters used 
$(C_i, \hbar\Omega^{H/L}_i, \sigma^i_{\rm ph})$ are
($0.2 \times 10^{-4}$, 327.2/326.6 meV, 20 meV) for $\nu$(OD),  ($1.0 \times 10^{-4}$, 301.2/301.8 meV, 8 meV) for $\nu$(OD-O), 
($1.2 \times 10^{-7}$, $76.8/76.5$ meV, 19 meV) for rot$_z$(OD-O), and $\Gamma^{H/L}_0=0.7 \times 10^2 /  1\times 10^{-4}$ s$^{-1}$. 
We first note that we need to use large values for the broadening $\sigma^i_\textrm{ph} \sim 10-20$ meV to fit 
the experimental data. We believe this to be caused by the statistical nature of the experimental data collection and 
do not reflect thermal or phonon DOS broadening\cite{Motobayashi}.
The fitting constants $C_i$ and $\Gamma^{H/L}_0$ are only determined up to a 
multiplicative factor since the occupations are determined by the ratios in Eq.~(\ref{eq.ratesol}).
We choose the prefactor of the shared $\nu$(OD-O) mode as the reference and set it to $10^{-4}$ based on an order of magnitude 
estimate from the experimental flip frequency\cite{Kumagai_2009}.
The values for $\Gamma^{H/L}_0$ only determine the low-bias occupation where $\Gamma^H_0 \gg \Gamma^L_0$
because the system is experimentally only observed in the $L$ configuration. The reason for this 
preference of $L$ at low bias might be due to the larger current and thus 
larger vibrational generation rate of low-energy phonons in the $H$ configuration. 
However, in the intermediate bias range $200\sim300$ meV we see a slight change in the occupation as shown in 
Fig.~\ref{fig.occupation}(b). Although we cannot unambiguously assign a vibrational mode
to this change in occupation from the available experimental data, we have chosen to model this by the 
rot$_z$(OD-O) mode. 
In contrast, there is much less latitude in the fitting parameters for the $\nu$(OD) and $\nu$(OD-O) modes.
We note that the change in occupation at 300 meV do not fit with the free $\nu$(OD) vibrational energy 
($\approx$ 330 meV) and clearly indicate that the shared  $\nu$(OD-O) mode 
is the main culprit in inducing the flip motion. This assignment is supported by the fact that 
the occupation quickly approaches 50/50 which implies that the emission rate constants 
$\lambda_\mathrm{em}$ of the high- and low-conductance 
states are of similar magnitude. This is clearly not the case for the free $\nu$(OD) but
true for the $\nu$(OD-O) mode, see Tab.~\ref{tab.hw} and Fig.~\ref{fig.emissionmap}.
To underline the effects of the shared $\nu$(OD-O) and free $\nu$(OD) stretch modes 
Fig.~\ref{fig.occupation} shows the full modeling (solid line) and without the free $\nu$(OD) vibration 
(dashed line). Without the $\nu$(OD) mode, the occupation rapidly approach the ratio of the 
shared $\nu$(OD-O) emission rate constants 
$n_L/n_H  \xrightarrow[eV \gg \hbar \omega]{} \lambda^H_\mathrm{em}/\lambda^L_\mathrm{em} \approx 0.71$. Note that the approximately 
equal occupation of the $H$/$L$ states is a coincidence and that the ratio of the occupations at high bias is 
simply given by the ratio of emission rate constants and the fitting constants $C_i$.
In contrast to the shared $\nu$(OD-O) mode, which increases the $H$ state occupation, the main effect of the 
free $\nu$(OD) mode is a slight increase of the $H\rightarrow L$ transition rate, Fig.~\ref{fig.occupation}(a), and thereby 
slightly decreases the occupation of the $H$ state.

\begin{figure}[t!]
\begin{center}
\includegraphics[width=\columnwidth]{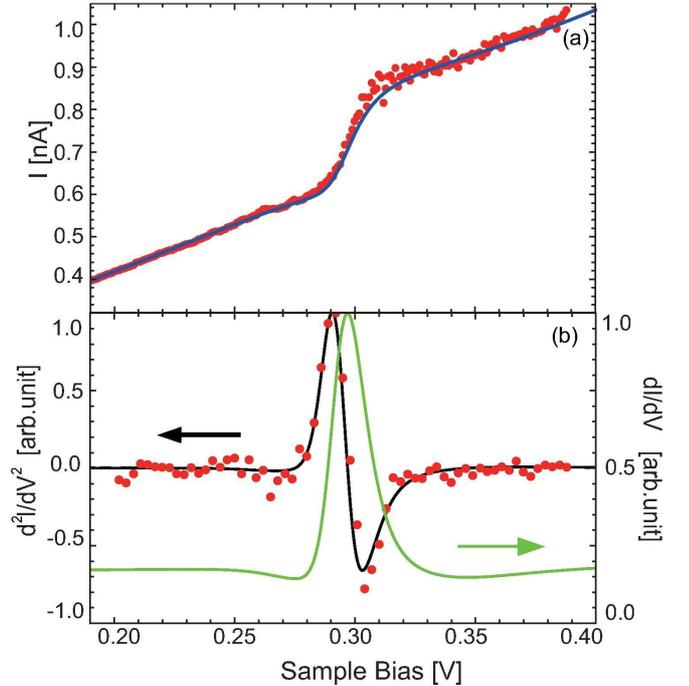}
\caption{(Color online) (a) $I-V$ curve (blue line: calculation, red dots: experiment) and (b) $dI/dV$ (green line: calculation) and $d^2I/dV^2$ (black line: calculation, red dots: experiment).
The experimental data originate from Ref.~\onlinecite{Kumagai_2009}.
\label{fig.IV}
}
\end{center}
\end{figure}

Using the calculated conductances ($\sigma_{H}=145.62$ nA/V, $\sigma_{L}=78.34$ nA/V)
scaled by the constant factor $2.56 \times 10^{-2}$ to account for the small tip-sample distance
used in the calculation and the bias dependent population [$n_{H}(V)$, $n_{L}(V)$] shown 
in Fig.~\ref{fig.occupation}(b), the $I-V$ characteristics were calculated from Eq.~(\ref{eq.IV}).
As shown in Fig.~\ref{fig.IV} the calculated time averaged $I-V$ curve [Fig.~4(a)] and $d^2I/dV^2$ [Fig.~4(b)] show
remarkable agreement with the experimental results.
We note that the lineshape of the $d^2I/dV^2$ signal is clearly different from inelastic electron 
tunneling spectra which normally only shows a peak or dip\cite{Paulsson2005}. In addition, the size of 
the signal from the vibration is much larger than what one normally associates with IETS spectra.

In summary, we have shown that the flip motion between high and low conductance 
configuration of the OD dimer on Cu(110) is mainly induced by the excitation of the hydrogen-bonded 
shared OD stretch mode. Because of the unique asymmetric inclined orientation of each 
hydroxyl, the shared and free OD stretch modes have different vibrational frequencies and 
consequently affect the flip motion at different applied bias voltages. 
The relative occupations ($n_H$ and $n_L$) of the high and low conductance 
state as a function of the bias voltage were nicely reproduced by solving a simple 
rate equation for $n_H$ and $n_L$ in terms of the transition rate between 
the $H$ and $L$ configurations. The calculated high and low conductance 
($\sigma_H$, $\sigma_L$) and their occupations [$n_H(V)$, $n_L(V)$] enabled us 
to obtain the nonlinear $I-V$ curve, $dI/dV$ and $d^2I/dV^2$ in excellent agreement 
with the experimental results. The presented theoretical analysis based on extensive DFT 
calculations (stable configurations, vibrational modes including their generation 
rates by tunneling electrons and damping rates) is not limited to the specific case of hydroxyl dimers 
on Cu(110), but can also be applied to other systems which exhibit nonlinear $I-V$
characteristics arising from the vibrationally mediated switching between high and low 
conductance states.

We thank H. Okuyama and T. Kumagai for valuable discussions. This work was supported by the 
Grant-in-Aid for Scientific Research B (No. 18340085) from the Japan Society 
for the Promotion of Science.



\end{document}